\documentclass[iop]{emulateapj}
\usepackage{epsfig,amsmath}
\usepackage[dvips]{color}
\usepackage{verbatim}
\usepackage{graphicx}
\usepackage{txfonts}
\usepackage{natbib}
\bibpunct{(}{)}{;}{a}{}{,}
\usepackage{hyperref}

\begin{document}

\title{Exploring Dust extinction at the edge of reionization\footnote{Based in part on observations done with the VLT/FORS2 (PI:081.A-0135(A)), VLT/ISAAC (programs:083.A-0552(A)) and VLT/SINFONI (083.A-0552(E)).}}

\author{Tayyaba~Zafar,\altaffilmark{1}
Darach~J.~Watson,\altaffilmark{1}
Nial~R.~Tanvir,\altaffilmark{2}
Johan~P.~U.~Fynbo,\altaffilmark{1}
Rhaana~L.~C.~Starling,\altaffilmark{2}
and~Andrew~J.~Levan\altaffilmark{3}}
\altaffiltext{1}{Dark Cosmology Centre, Niels Bohr Institute, University of Copenhagen,
Juliane Maries Vej 30, DK-2100 Copenhagen \O, Denmark; tayyaba, darach, jfynbo@dark-cosmology.dk}
\altaffiltext{2}{Department of Physics and Astronomy, University of Leicester, University Road Leicester, LE1 7RH, UK; nrt3, rlcs1@star.le.ac.uk}
\altaffiltext{3}{Department of Physics, University of Warwick, Coventry CV4 7AL, UK; A.J.Levan@warwick.ac.uk}
\begin{abstract}
The brightness of gamma-ray burst (GRB) afterglows and their occurrence in young,
blue galaxies make them excellent probes to study star forming regions
in the distant Universe. We here elucidate dust extinction properties in the early
Universe through the analysis of the afterglows of all known $z>6$ GRBs: GRB\,090423, 080913
and 050904, at $z=8.2,~6.69$, and $6.295$, respectively.
We gather all available optical and near-infrared photometry,
spectroscopy and X-ray data to construct spectral energy distributions (SEDs) at
multiple epochs. We then fit the SEDs at all epochs with a dust-attenuated 
power-law or broken power-law. We find no evidence for 
dust extinction in GRB\,050904 and GRB\,090423, with possible evidence for a low
level of extinction in GRB\,080913. We compare the high redshift GRBs to a sample of lower
redshift GRB extinctions and find a lack of even moderately extinguished events
($A_V\sim0.3$) above $z\gtrsim4$. In spite of the biased selection and small number
statistics, this result hints at a decrease in dust content in star-forming
environments at high redshifts.

\end{abstract}

\maketitle

%
\section{Introduction\label{introduction}}
Dust formation in the early universe is a hotly debated topic. The properties and quantity
of dust at these times are essential not only to our understanding of how dust forms and
evolves, but also to models of star-formation and the appearance of the first 
generations of stars and galaxies. Currently we extrapolate dust properties in the local 
group to very different environments that existed less than a billion years after the Big
Bang. Long-duration gamma-ray bursts (GRBs), associated with the deaths of massive
stars \citep{woosley, galama, hjorth, stanek, malesani,campana06}, have extremely bright and spectrally simple afterglows $(F_{\nu}\propto 
\nu^{-\beta})$ \citep{sari, granot}, and so provide an opportunity to obtain not only
effective reddening, but also absolute extinctions from distant galaxies (e.g. \citealt{watson06}; \citealt{schady}; \citealt{zafar11}). The recent tremendous advances
in this subject were enabled by the dedicated GRB satellite, \emph{Swift} \citep{gehrels},
with its  precise localization and fast responding X-ray and UV/optical telescopes, 
detecting the afterglow within only a few tens of seconds after the
$\gamma$-ray trigger. The data from \emph{Swift} have provided a very large sample of
afterglows virtually complete in X-ray detections, and with a high completeness in the
optical. It has also for the first time, enabled the discovery of GRBs at redshifts greater
than or comparable to the most distant known galaxies and quasars (QSO) \citep[see][]{kawai,greiner,tanvir,salvaterra}.
 
In this paper we collect for the first time all the available afterglow photometry
and spectra from near-infrared (NIR) to X-ray at different epochs for all GRBs at
$z>6$: GRB\,050904 ($z=6.295$), GRB\,080913 ($z=6.69$) and GRB\,090423 
($z\sim8.2$). Since observations of these high-z GRBs are being made in the rest frame
UV, they are in principle relatively sensitive to even small amounts of dust. We extract the SEDs at multiple epochs for all three GRBs and jointly fit the X-ray to NIR data to determine the properties of dust at $z>6$. 
In \S2 we describe multi-wavelength observations of the afterglows at
different epochs carried out with different instruments. In \S3 we present our
results from the SED fitting. Based on our results, in \S4 we discuss the origin of
dust at high redshift and compare to the first large spectroscopic sample of
GRB-derived extinction curves, the first spectroscopic sample of absolute
extinction curves outside the local group. In \S5 we provide our conclusions. 

For all parameters we quote uncertainties at the 68\% confidence level. We
derive $3\sigma$ upper limits for the cases where detection is less than
$2\sigma$ significant. A cosmology where $H_0=72$\,km\,s$^{-1}$\,Mpc$^{-1}$,
$\Omega_\Lambda = 0.73$ and $\Omega_{\rm m}=0.27$ is assumed throughout.

%
%
\section{Multi-wavelength data}\label{Multi-wavelength data} 
The X-ray to NIR SED of GRB\,050904 was analysed in detail in a previous work \citep[see][]{zafar}. 

On 2008 September 13 at 06:48:33.6 UT \emph{Swift}-XRT began observing GRB\,080913. Telescopes at the European Southern Observatory (ESO), Gemini Observatories, and the Subaru all observed the field of GRB\,080913 in different optical and NIR bands \citep{greiner}. On 2008 September 13, an optical spectrum of the afterglow of GRB\,080913 was secured at the Very Large Telescope (VLT) using the FOcal Reducer and low dispersion Spectrograph 2 (FORS2) \citep{greiner, fynbo, patel}. The spectrum shows a sharp break at $\sim9400\,\AA$ supporting the detection of Ly$\alpha$ absorption at $z=6.69$. We use this spectrum in addition to photometric data in the $J$, $H$ and $K$ bands from the lightcurves presented in \citet{greiner}. We corrected the data for extinction in the Milky Way (MW) of $E(B-V) = 0.043$\,mag along the line of sight to this burst \citep{schlegel}.

At 07:56:32 UT on 2009 April 23, \emph{Swift}-XRT began observations of GRB\,090423. The afterglow of GRB\,090423 was observed in optical and NIR bands, with non-detection in $griz^\prime$ and $Y$-band due to the high redshift \citep{tanvir}. A NIR spectrum was obtained with the VLT Infrared Spectrometer And Array Camera (ISAAC) starting about 17.5 hours after the burst trigger. A redshift of $\sim8.2$ was estimated from a break observed around $1.14\,\mu$m due to Ly$\alpha$ absorption by the neutral intergalactic medium (IGM). Another spectrum of the afterglow was obtained 40 hours after the burst with the VLT's Spectrograph using INtegral Field Observations in the Near Infrared (SINFONI) and confirms the analysis of the ISAAC spectrum \citep{tanvir}. The finding was also confirmed with a spectrum obtained with Telescopio Nazionale Galileo (TNG) at $\sim14$ hours after the burst \citep{salvaterra}. We use the $J$, $H$ and $K$ band photometry of \citet{tanvir}. The data were corrected for Galactic extinction of $E(B-V)=0.029$\,mag \citep{schlegel}.

\begin{table}
\begin{minipage}[t]{\columnwidth}
\caption{Best fit parameters of the afterglow SEDs at different epochs with \textbf{Small Magellanic Cloud (SMC)} type extinction. The provided upper limits are $3\sigma$.}      
\label{table:1} 
\centering
\begin{tabular}{@{}c c c c c c@{}}   
\hline\hline                        
GRB & Epoch & $A_V$ & $\beta_1$ & $\beta_2$ & log $\nu_{break}$\\
 &  & (mag)  &  & & Hz \\ 
\hline\hline
050904 & 0.47d & $<0.17$ & $1.23 \pm 0.08$ & \mbox{\ldots} &  \mbox{\ldots} \\
	& 1.25d & $<0.08$ &  $1.17 \pm 0.51$ &  \mbox{\ldots} &  \mbox{\ldots} \\
	& 3.4d & $<0.16$ & $1.24 \pm 0.07$ &  \mbox{\ldots} &  \mbox{\ldots} \\[5pt]
\hline
080913 & 10.5m & $0.14 \pm 0.06$ & $ 0.78 \pm 0.02$ & \mbox{\ldots} &  \mbox{\ldots} \\
	& 26.9m & $0.22 \pm 0.11$ &  $ 0.65 \pm 0.04$ &  \mbox{\ldots} &  \mbox{\ldots} \\
	& 1.88hr & $0.12 \pm 0.03$ & $ 0.79 \pm 0.03$ &  \mbox{\ldots} &  \mbox{\ldots} \\
	& 4.78d & $<0.40$ & $ 0.94 \pm 0.04$ &  \mbox{\ldots} &  \mbox{\ldots} \\[5pt]
\hline
090423 & 16.7hr & $<0.10$ & $ 0.45 \pm 0.11$ & $0.95\pm0.11$ & $15.1\pm0.7$ \\
	& 1.65d & $<0.09$ &  $0.68\pm0.13$ & $1.18\pm0.13$ &  $16.5\pm0.9$ \\[5pt]
\hline
\end{tabular}
\end{minipage}
\end{table}

For all three GRB afterglows the \emph{Swift} XRT data were available and downloaded from the \emph{Swift} data archive and reduced using HEAsoft (version 6.10). Spectra and lightcurves were extracted in the 0.3--10.0 keV energy range. The X-ray spectra at different epochs were analyzed using the latest associated calibration files. All X-ray spectra were obtained near the time of optical/NIR spectra or relevant available photometry. To get an approximate X-ray flux level at each epoch, the afterglow lightcurves were obtained from the \emph{Swift} XRT GRB light curve repository at the UK Swift Science Data Centre\footnote[1]{\url{www.swift.ac.uk/xrt_curves}}, created as described in \citet{evans,evans10} and fitted by assuming a smoothly decay broken power-law \citep{beuermann}. Considering the photon weighted mean time, the X-ray spectra were then normalized to the relevant SED time by using the lightcurve fit. 

The XMM-\emph{Newton} observations of the afterglows of GRB\,080913 and GRB\,090423 were performed at $\sim4.4$ and $\sim2.2$ days respectively. The data were collected from the XMM-\emph{Newton} science archive (XSA)\footnote[2]{\url{http://xmm.esac.esa.int/xsa/}}. The data were reduced in 0.3--10.0 keV band using Science Analysis System (SAS, version 10.0.0). The spectrum of GRB\,080913 was fitted using a single power-law with a best fit photon index of $\Gamma=2.4\pm0.4$ and frozen for Galactic X-ray absorption of $3.2\times10^{20}$ cm$^{-2}$ \citep[using the nH FTOOL;][]{kalberla}. The best fit photon index for GRB\,090423 is $\Gamma=2.4\pm0.3$ with the fixed Galactic absorption of $2.9\times10^{20}$ cm$^{-2}$ \citep{kalberla}.

Parts of spectra with a restframe wavelength shortward of Ly$\alpha$ (i.e.\ $\lambda<1216\AA$) are not included in the SED analysis to avoid absorption caused by the neutral IGM. The absorption lines arising from ionic species were removed from the spectrum of GRB\,050904. Uncertainty in the Galactic extinction does not affect our results. The \citet{dutra} reddening maps confirm the \citet{schlegel} Galactic extinction maps up to $E(B-V)=0.25$\,mag. Assuming an uncertainty of 15\% \citep{schlafly} even for the case with the  largest Galactic extinction ($z^\prime$-band, GRB\,050904), this corresponds to an uncertainty in the Galactic reddening value of $\lesssim0.015$\, mag. This indicates that the uncertainty due to the Galactic extinction correction in the NIR is always smaller than our statistical uncertainties.  
%
%
\section{Results}\label{results}

SEDs of the three GRB afterglows were obtained at multiple epochs. The data were fitted with dust-attenuated power-laws or broken power-laws. In addition to our standard extinction model \citep[SMC type, see][]{schady,zafar11}, we also fitted the data with the extinction curve inferred by  \citet{maiolino} to explain the spectrum of a QSO at $z=6.2$ \citep[although see][]{gallerani2}. 

\begin{figure}
  \centering
   {\includegraphics[width=\columnwidth,clip=]{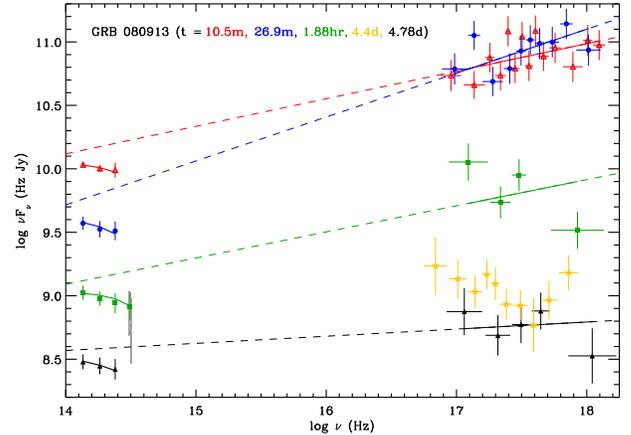}}
     \caption{NIR to X-ray SED of the afterglow of GRB\,080913 at 10.5 minutes (open red triangles), 26.85 minutes (blue circles), 1.88 hours (green squares) and 4.78 days (black triangles). Yellow stars correspond to XMM-\emph{Newton} data at 4.4 days after the burst. The grey curve is the optical spectrum of the afterglow taken at $t_0+1.88$ hours. The solid lines represent the best-fit to the data; dashed lines illustrate the unextinguished power-law from the best fit. Being close in time, the X-ray spectra for the first two epochs are superimposed.\\}
        \label{080913}
   \end{figure}

\subsection{Multi epoch SED of GRB\,050904}
The SED of this burst was examined in detail in \citet{zafar}, where the SED of the afterglow of GRB\,050904 is constructed at 0.47, 1.25 and 3.4 days after the burst. We use the results from the analysis here. The results of the SMC-extinction fits are reported in Table.~\ref{table:1} for completeness. The fits using supernova (SN)-origin extinction from \citet{maiolino} suggest that there is no evidence in this burst for this type of extinction. The SED of GRB\,050904 has been fitted by \citet{haislip, tagliaferri, kann10} finding no dust extinction.

\subsection{Multi epoch SED of GRB\,080913}
   
The SED of the afterglow of GRB\,080913 is extracted at four epochs i.e.\ 10.5 and 26.9~minutes, 1.88~hours and 4.78~days after the burst (see Fig.~\ref{080913}). To get precise photometry in the $z$-band which is affected by strong Ly$\alpha$ absorption, we performed spectro-photometric analysis at 1.88 hours using the effective transmission function of the FORS2 $z$-Gunn filter. We constructed SEDs of the stars in the field using the Sloan Digital Sky Survey \citep[SDSS;][]{fukugita} and Two Micron All Sky Survey \citep[2MASS;][]{skrutskie} catalogues. We derived the count-to-flux conversion factor and estimated the afterglow band-integrated flux from its measured counts. We computed the flux density at $\lambda_{\rm{rest}}=1250\,\AA$ using the procedure described in \citet{zafar}. At 1.88~hour the flux density at $1250\,\AA$ was $2.66\pm0.63~\mu$Jy. The NIR to X-ray SED at all epochs is well fitted with a single power-law and SMC type extinction ($\chi^2\rm{/dof}=47/81$). The combined fit for dust extinction at all epochs resulted in marginal dust reddening of $A_V=0.12\pm0.03$. Fitting each epoch individually, we find dust extinction significant at the $2\sigma$ level at 10.5 and 26.9 min, at $4\sigma$  at 1.88 hours and $1\sigma$ at 4.78 days (see Table~\ref{table:1}). We also fitted the data using the \citet{maiolino} extinction curve, resulting in a significantly worse fit ($\chi^2\rm{/dof}=69/81$) and zero extinction. We also fitted the SED with Milky Way and Large Magellanic Cloud
extinction curves but found no statistically significant improvement in the fit compared to the SMC, and we found no evidence for a 2175\,\AA\ bump. As the simpler model, we used the SMC to parameterize
the extinction here and to parameterize the upper limit in GRB\,090423 below.

\begin{figure}
  \centering
   {\includegraphics[width=\columnwidth,clip=]{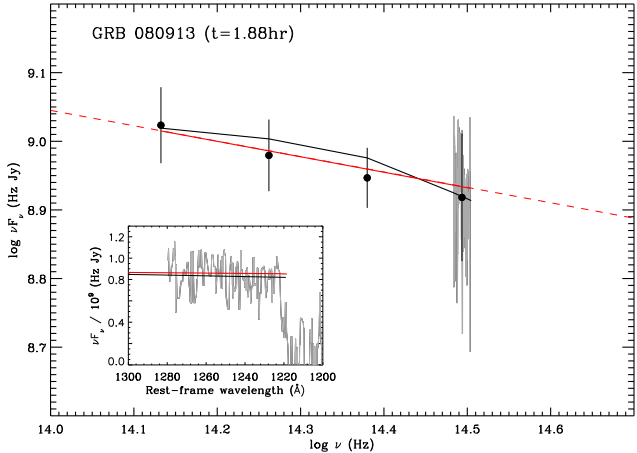}}
     \caption{NIR and optical SED of the afterglow of GRB\,080913 at 1.88 hours after the trigger. The grey curve represents the optical spectrum. The solid black curve corresponds to the best fit SMC-dust--attenuated power-law to the full NIR--X-ray SED, while the red line shows the best fit power-law to the NIR-optical data only with no extinction. \emph{Inset:} Zoom of the optical spectrum and fits. The spectrum has been median filtered for display purposes.\\}
        \label{spec}
   \end{figure}

It is worth noting that the X-ray lightcurve for the afterglow of GRB\,080913 shows some evidence of flaring, possibly resulting in harder spectra \citep[see][]{evans,greiner}. If the SED of the afterglow of GRB\,080913 is dominated by two separate components in optical-NIR and X-ray wavelengths, as observed at early times for GRB\,050904, we should fit the optical-NIR data independent of the X-rays. Using this method, we obtain results consistent with a single power-law and no dust extinction, with a spectral slope of $\beta=1.2\pm0.1$  and an upper limit of $A_V<0.09$ (see Fig.~\ref{spec}). Previously \citet{greiner,kann10} fitted the optical-NIR SED of GRB 080913 and found no dust extinction.

\subsection{Multi epoch SED of GRB\,090423}
\begin{figure}
  \centering
   {\includegraphics[width=\columnwidth,clip=]{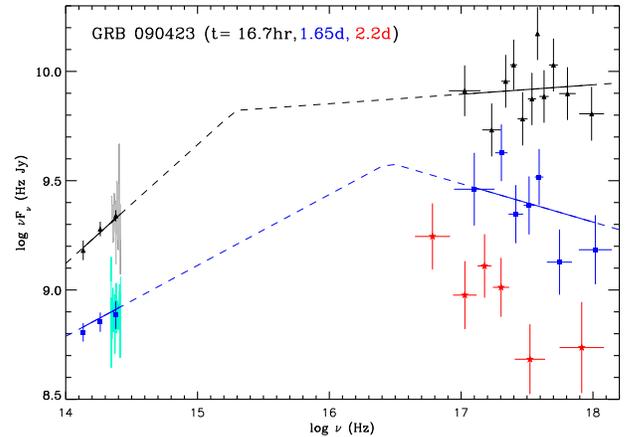}}
     \caption{The NIR to X-ray SED of the afterglow of GRB\,090423 at 16.7 hours (black triangles) and 1.645 days (blue squares). The corresponding lines show the best fits to the data. The ISAAC and SINFONI spectra are shown with the grey and cyan curves respectively. Red stars represent the XMM-\emph{Newton} data taken at 2.21 days after the trigger.\\}
        \label{090423}
   \end{figure}

The SED of the afterglow of GRB\,090423 was constructed at 16.7~hours and 1.654~days (see Fig.~\ref{090423}). The VLT/ISAAC and SINFONI spectra, NIR photometry and X-ray spectra were fitted at both epochs. The data from both epochs were well fit with no dust using a broken power-law. The estimated $3\sigma$ upper limit on the $A_V$ with SMC extinction is 0.10 and 0.09\,mag at 16.7 hours and 1.654 epochs respectively. 

%
%

\section{Discussion}\label{discussion}

Two apparently conflicting themes appear in studies of dust at high redshift. The first is the expected decline in extinction as we get to early in the Universe, where galaxies are younger and dust has less time to form. In the local Universe low and intermediate mass asymptotic giant branch (AGB) stars are believed to be the principal sources of interstellar dust, requiring $\gtrsim 1$\,Gyr to evolve off the main sequence to produce dust \citep{gehrz, morgan,marchenko,dwek}. Because of the time required to evolve to the giant branch, it has been posited that AGB stars should not contribute significantly to dust at $z\gtrsim 6$ \citep{dwek}. Spectroscopic \citep{douglas} and photometric \citep{verma} studies of $z\sim5$ lyman-break galaxies seem to confirm the general trend of little dust at high redshifts, showing low extinctions, and the sample of $z$-band dropout galaxies ($z\sim7$) shows little or no dust \citep{valentino}.

However, the second, at first paradoxical, theme is the very large dust masses discovered in quasar host galaxies at $z\sim6$  from mm and sub-mm observations \citep[e.g.][]{bertoldi03,carilli,beelen,wang,michalowski,gall}, and the mystery surrounding the origin of such large quantities of dust within a billion years of the Big Bang. It has been suggested that dust produced on short time scales could be produced in the ejecta of Type II SNe \citep{kozasa,todini, morgan, nozawa, dwek, marchenko, hirashita}, although observations of local SNe do not support this hypothesis \citep[e.g.][]{meikle,smith}. \citet{valiante} estimated that under certain conditions AGB stars could produce dust at earlier epochs, beginning to dominate between 150 and 500\,Myr. However \citet{michalowski} found that given the stellar masses in high redshift QSO hosts, AGB stars could not be responsible for the large masses of dust observed in these objects at $z>5$ \citep[see][]{cherchneff,pipino,gall10}, and suggested rapid dust growth in the ISM as an alternative \citep[see also][]{draine}. The modeling of Lyman alpha emitters at $z\sim6$ suggests that the largest galaxies are the most dust enriched and that dust content decreases with increasing redshift \citep{lai,finkelstein09,dayal10}. In spite of the emphasis in the literature on these QSO hosts, it may be that such very massive systems, being exceptional and rare, tell us little about typical star forming environments at very high redshift. And while, young star-forming galaxies at high redshift with strong nebular emission appear to be somewhat dust-obscured in spite of their blue continuua \citep{schaerer,watson10}, the quantities of dust are not large, consistent with the idea that in the early universe, most of the star-formation is likely to be less obscured than at later times. GRBs, associated as they are with the deaths of massive stars, usually occur in young, blue and sub-luminous galaxies with a high specific star formation rate \citep{lefloch,courty,christensen,prochaska,savaglio09,castro-ceron} and are therefore likely to be good probes of star-forming regions, especially at high redshift.

\begin{figure}
  \centering
   {\includegraphics[width=\columnwidth,clip=]{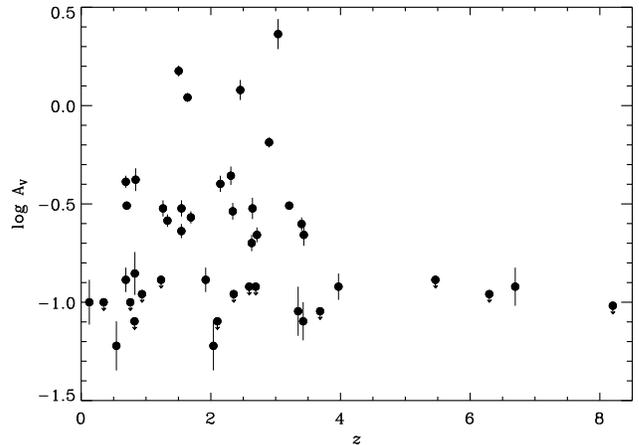}}
     \caption{Extinction in GRB afterglow versus redshift. The three high redshift GRBs ($z>6$) are from this work while the 42 low redshift GRBs are from a sample studied in \citet{zafar11}. There seems to be a dearth of GRB afterglows at $z\gtrsim4$ with $A_V\sim0.3$.\\}
        \label{zplot}
   \end{figure}

In this work, we have determined that all the identified $z>6$ GRBs have low or negligible extinctions. By comparing these results with the first spectroscopic GRB extinction sample of \citet{zafar11}, we find that at high-redshifts ($z\gtrsim4$), GRB sightlines appear significantly less extinguished (see Fig. \ref{zplot}): the majority of reddened low-redshift GRBs are observed with $A_V\sim0.3$\,mag. While the numbers are small, the absence of any GRBs at $z\gtrsim4$ with extinctions in this range is striking. The first explanation to examine is observational bias, where higher-redshift GRBs will be observed farther into the restframe UV where the extinction is likely to be more severe, possibly preventing spectra from being obtained or the afterglow from being detected. For the $z>6$ GRBs, a restframe $A_V=0.3$ corresponds to 1--1.5 magnitudes of extinction in the observed $J$-band and 0.5--0.8 magnitudes in the observed $K$, for an SMC extinction curve. Considering the photometric observations given in \citet{ruiz07,haislip,tagliaferri,greiner,tanvir}, we found that even with $A_V=0.3$ the GRBs at $z>6$ could have been detected with relative ease and a photometric redshift obtained. In the case of GRB\,050904, it seems likely that a spectroscopic redshift would have been obtained even with restframe $A_V=0.3$. This work therefore hints at a low extinction of most $z>4$ bursts, since it seems that our $z>6$ bursts would have been found even with $A_V=0.3$. Therefore we have looked at all known $z>5$ GRB afterglows in the literature. There are four bursts with firm estimates of $z>5$ known. Three of them have estimates of their dust extinction available. The four bursts are: GRB\,050814 \citep{jakobsson06} at $z\sim5.3$, GRB\,060522 \citep{cenko06} at $z=5.11$, GRB\,060927 (\citealt{ruiz07}; \citealt{zafar11}) at $z=5.46$ and GRB\,071025 \citep{perley} at $z\sim5$. \citet{jakobsson06} obtained a photometric redshift of the afterglow of GRB\,050814 and the SED is consistent with no dust reddening (P.~Jakobsson, private comm. 2010). \citep{cenko06} give a spectroscopic redshift for GRB\,060522 based on the Ly$\alpha$ break, however no estimate of the extinction is available so far in the literature. \citet{ruiz07} obtained the spectroscopic redshift of the afterglow of GRB\,060927. We have studied the SED of the afterglow of this burst in our spectroscopic sample of \citet{zafar11} and it is consistent with $A_V=0$ (see Fig. \ref{zplot}). Moderate extinction is claimed in the case of GRB\,071025 \citep{perley}, with extinction corresponding to $A_V\sim0.5$ (converting $A_{3000\,\AA}$ to $A_V$ using the SMC extinction curve ratio). A \citet{maiolino} extinction curve is required to fit the SED. With this extinction, GRB\,071025 has a high enough extinction to begin to fill in the gap at high redshift around $A_V\sim0.3$, though it is perhaps notable that it has the lowest redshift of all the bursts examined here. It is more anomalous in the sense that it has an apparently unique extinction curve since reanalyses have shown that no other QSO or GRB any longer require such a peculiar extinction \citep[see][]{gallerani,zafar}.

The issues of bias and low number statistics are clearly important and we cannot resolve them with this sample. The fact that we infer very little dust in the afterglows studied here does not imply that dusty environments do not exist along sightlines to $z>6$ GRBs, but it suggests that there is less dust in highly star-forming regions at $z\gtrsim4$. This result is also potentially interesting in the context of dark bursts -- suggesting that highly-extinguished bursts may not exist at high redshift, and that bright, very high redshift GRBs will be detectable as long as they are observed in the NIR.

Finally, the comparison to the column densities of metals is striking. In the cases of both GRB\,050904 and GRB\,090423, column densities of $N_{H,X }\sim3\times10^{22}$ and $1\times10^{23}$ are claimed (assuming solar abundances and neutral elements, \citealt{watson062}; \citealt{campana}; \citealt{gendre}; \citealt{tanvir}; Starling et al. in prep. 2011) from the soft X-ray absorption. The dust-to-metals ratios in both cases are at most a few percent of the value in the local group. Given that the metals are likely to be highly-ionized \citep{schady,watson07}, the metal column must be substantially higher than found using neutral
medium models. This could drive the dust-to-metals ratio down below a percent of local group values. While it is established that GRB environments can have low dust-to-metals ratios \citep{ardis, watson06, schady}, these high-redshift objects are extreme. Such extreme ratios could be related to the more effective destruction of dust due to the high luminosity of these bursts \citep[e.g.][]{fruchter}; the effects of dust destruction may be
discernible not only in very early colour changes in the GRB prompt phase UV emission, but also in the extinction curve at later times \citep{perna02}. A small dust-to-metals ratio could however, also be due to something more fundamental, such as the rapid production of metals in SNe but delayed formation and growth of the dust. For the metals to form before the dust suggests at least several million years between the formation of metals and the onset of significant dust growth. Observations of a larger sample may help to clarify this issue (see \citealt{zafar11}).

%
%
\section{Conclusions\label{conclusions}}
We have investigated cosmic dust at $z>6$ using the afterglows of the highest redshift GRBs known. We performed multi-epoch NIR--X-ray SED analysis of GRB\,050904, 080913, and 090423. We infer from our analysis that there is no evidence of dust in GRB\,050904 and GRB\,090423. We find possible evidence of low extinction in GRB\,080913 consistent with an SMC extinction curve, however the optical spectrum and NIR photometry alone are consistent with a power-law with no dust extinction. In no case did we find evidence for any other extinction curve (e.g.\ the `SN-origin' curve of \citealt{maiolino}). Comparing to a much larger spectroscopic sample of GRB extinctions, we find a distinct absence of extinction at high redshifts. While the high redshift sample is very small and some of the effect may be explained by restframe UV-selection bias, this is not the whole picture, as all the $z>6$ GRBs would have been detected at the typical $A_V$ of lower redshift reddened bursts. This hints that there is less dust along sightlines to the highest redshift GRBs, indicating less dust in the early Universe. From these results, we also infer an extremely low dust-to-metals ratio in GRBs at high redshift, suggestive either of efficient dust destruction, or a delay of at least several million years between the formation of metals and the formation and growth of dust.

\begin{acknowledgements} 
The Dark Cosmology Centre is funded by the Danish National Research Foundation. We are grateful to Jens Hjorth, Daniele Malesani, Johan Richard and Sune Toft for helpful discussion and comments.
\end{acknowledgements}

\bibliography{high_z.bib}

\begin{thebibliography}{80}
\expandafter\ifx\csname natexlab\endcsname\relax\def\natexlab#1{#1}\fi

\bibitem[{{Beelen} {et~al.}(2006){Beelen}, {Cox}, {Benford}, {Dowell},
  {Kov{\'a}cs}, {Bertoldi}, {Omont}, \& {Carilli}}]{beelen}
{Beelen}, A., {Cox}, P., {Benford}, D.~J., {Dowell}, C.~D., {Kov{\'a}cs}, A.,
  {Bertoldi}, F., {Omont}, A., \& {Carilli}, C.~L. 2006, \apj, 642, 694

\bibitem[{{Bertoldi} {et~al.}(2003){Bertoldi}, {Carilli}, {Cox}, {Fan},
  {Strauss}, {Beelen}, {Omont}, \& {Zylka}}]{bertoldi03}
{Bertoldi}, F., {Carilli}, C.~L., {Cox}, P., {Fan}, X., {Strauss}, M.~A.,
  {Beelen}, A., {Omont}, A., \& {Zylka}, R. 2003, \aap, 406, L55

\bibitem[{{Beuermann} {et~al.}(1999){Beuermann}, {Hessman}, {Reinsch},
  {Nicklas}, {Vreeswijk}, {Galama}, {Rol}, {van Paradijs}, {Kouveliotou},
  {Frontera}, {Masetti}, {Palazzi}, \& {Pian}}]{beuermann}
{Beuermann}, K., {et~al.} 1999, \aap, 352, L26

\bibitem[{{Campana} {et~al.}(2006){Campana}, {Mangano}, {Blustin}, {Brown},
  {Burrows}, {Chincarini}, {Cummings}, {Cusumano}, {Della Valle}, {Malesani},
  {M{\'e}sz{\'a}ros}, {Nousek}, {Page}, {Sakamoto}, {Waxman}, {Zhang}, {Dai},
  {Gehrels}, {Immler}, {Marshall}, {Mason}, {Moretti}, {O'Brien}, {Osborne},
  {Page}, {Romano}, {Roming}, {Tagliaferri}, {Cominsky}, {Giommi}, {Godet},
  {Kennea}, {Krimm}, {Angelini}, {Barthelmy}, {Boyd}, {Palmer}, {Wells}, \&
  {White}}]{campana06}
{Campana}, S., {et~al.} 2006, \nat, 442, 1008

\bibitem[{{Campana} {et~al.}(2011){Campana}, {Salvaterra}, {Tagliaferri},
  {Kouveliotou}, \& {Grindlay}}]{campana}
{Campana}, S., {Salvaterra}, R., {Tagliaferri}, G., {Kouveliotou}, C., \&
  {Grindlay}, J. 2011, \mnras, 410, 1611

\bibitem[{{Carilli} {et~al.}(2001){Carilli}, {Bertoldi}, {Omont}, {Cox},
  {McMahon}, \& {Isaak}}]{carilli}
{Carilli}, C.~L., {Bertoldi}, F., {Omont}, A., {Cox}, P., {McMahon}, R.~G., \&
  {Isaak}, K.~G. 2001, \aj, 122, 1679

\bibitem[{{Castro Cer{\'o}n} {et~al.}(2010){Castro Cer{\'o}n},
  {Micha{\l}owski}, {Hjorth}, {Malesani}, {Gorosabel}, {Watson}, {Fynbo}, \&
  {Morales Calder{\'o}n}}]{castro-ceron}
{Castro Cer{\'o}n}, J.~M., {Micha{\l}owski}, M.~J., {Hjorth}, J., {Malesani},
  D., {Gorosabel}, J., {Watson}, D., {Fynbo}, J.~P.~U., \& {Morales
  Calder{\'o}n}, M. 2010, \apj, 721, 1919

\bibitem[{{Cenko} {et~al.}(2006){Cenko}, {Berger}, {Djorgovski}, {Mahabal}, \&
  {Fox}}]{cenko06}
{Cenko}, S.~B., {Berger}, E., {Djorgovski}, S.~G., {Mahabal}, A.~A., \& {Fox},
  D.~B. 2006, GRB Coordinates Network, 5155, 1

\bibitem[{{Cherchneff} \& {Dwek}(2009)}]{cherchneff}
{Cherchneff}, I., \& {Dwek}, E. 2009, \apj, 703, 642

\bibitem[{{Christensen} {et~al.}(2004){Christensen}, {Hjorth}, \&
  {Gorosabel}}]{christensen}
{Christensen}, L., {Hjorth}, J., \& {Gorosabel}, J. 2004, \aap, 425, 913

\bibitem[{{Courty} {et~al.}(2004){Courty}, {Bj{\"o}rnsson}, \&
  {Gudmundsson}}]{courty}
{Courty}, S., {Bj{\"o}rnsson}, G., \& {Gudmundsson}, E.~H. 2004, \mnras, 354,
  581

\bibitem[{{Dayal} {et~al.}(2010){Dayal}, {Ferrara}, \& {Saro}}]{dayal10}
{Dayal}, P., {Ferrara}, A., \& {Saro}, A. 2010, \mnras, 402, 1449

\bibitem[{{Douglas} {et~al.}(2010){Douglas}, {Bremer}, {Lehnert}, {Stanway}, \&
  {Milvang-Jensen}}]{douglas}
{Douglas}, L.~S., {Bremer}, M.~N., {Lehnert}, M.~D., {Stanway}, E.~R., \&
  {Milvang-Jensen}, B. 2010, \mnras, 409, 1155

\bibitem[{{Draine}(2009)}]{draine}
{Draine}, B.~T. 2009, in Astronomical Society of the Pacific Conference Series,
  Vol. 414, Cosmic Dust - Near and Far, ed. {T.~Henning, E.~Gr{\"u}n, \&
  J.~Steinacker}, 453

\bibitem[{{Dutra} {et~al.}(2003){Dutra}, {Ahumada}, {Clari{\'a}}, {Bica}, \&
  {Barbuy}}]{dutra}
{Dutra}, C.~M., {Ahumada}, A.~V., {Clari{\'a}}, J.~J., {Bica}, E., \& {Barbuy},
  B. 2003, \aap, 408, 287

\bibitem[{{Dwek} {et~al.}(2007){Dwek}, {Galliano}, \& {Jones}}]{dwek}
{Dwek}, E., {Galliano}, F., \& {Jones}, A.~P. 2007, \apj, 662, 927

\bibitem[{{El{\'{\i}}asd{\'o}ttir} {et~al.}(2009){El{\'{\i}}asd{\'o}ttir},
  {Fynbo}, {Hjorth}, {Ledoux}, {Watson}, {Andersen}, {Malesani}, {Vreeswijk},
  {Prochaska}, {Sollerman}, \& {Jaunsen}}]{ardis}
{El{\'{\i}}asd{\'o}ttir}, {\'A}., {et~al.} 2009, \apj, 697, 1725

\bibitem[{{Evans} {et~al.}(2007){Evans}, {Beardmore}, {Page}, {Tyler},
  {Osborne}, {Goad}, {O'Brien}, {Vetere}, {Racusin}, {Morris}, {Burrows},
  {Capalbi}, {Perri}, {Gehrels}, \& {Romano}}]{evans}
{Evans}, P.~A., {et~al.} 2007, \aap, 469, 379

\bibitem[{{Evans} {et~al.}(2010){Evans}, {Willingale}, {Osborne}, {O'Brien},
  {Page}, {Markwardt}, {Barthelmy}, {Beardmore}, {Burrows}, {Pagani},
  {Starling}, {Gehrels}, \& {Romano}}]{evans10}
---. 2010, \aap, 519, A102

\bibitem[{{Finkelstein} {et~al.}(2009){Finkelstein}, {Rhoads}, {Malhotra}, \&
  {Grogin}}]{finkelstein09}
{Finkelstein}, S.~L., {Rhoads}, J.~E., {Malhotra}, S., \& {Grogin}, N. 2009,
  \apj, 691, 465

\bibitem[{{Fruchter} {et~al.}(2001){Fruchter}, {Krolik}, \&
  {Rhoads}}]{fruchter}
{Fruchter}, A., {Krolik}, J.~H., \& {Rhoads}, J.~E. 2001, \apj, 563, 597

\bibitem[{{Fukugita} {et~al.}(1996){Fukugita}, {Ichikawa}, {Gunn}, {Doi},
  {Shimasaku}, \& {Schneider}}]{fukugita}
{Fukugita}, M., {Ichikawa}, T., {Gunn}, J.~E., {Doi}, M., {Shimasaku}, K., \&
  {Schneider}, D.~P. 1996, \aj, 111, 1748

\bibitem[{{Fynbo} {et~al.}(2009){Fynbo}, {Jakobsson}, {Prochaska}, {Malesani},
  {Ledoux}, {de Ugarte Postigo}, {Nardini}, {Vreeswijk}, {Wiersema}, {Hjorth},
  {Sollerman}, {Chen}, {Th{\"o}ne}, {Bj{\"o}rnsson}, {Bloom}, {Castro-Tirado},
  {Christensen}, {De Cia}, {Fruchter}, {Gorosabel}, {Graham}, {Jaunsen},
  {Jensen}, {Kann}, {Kouveliotou}, {Levan}, {Maund}, {Masetti},
  {Milvang-Jensen}, {Palazzi}, {Perley}, {Pian}, {Rol}, {Schady}, {Starling},
  {Tanvir}, {Watson}, {Xu}, {Augusteijn}, {Grundahl}, {Telting}, \&
  {Quirion}}]{fynbo}
{Fynbo}, J.~P.~U., {et~al.} 2009, \apjs, 185, 526

\bibitem[{{Galama} {et~al.}(1998){Galama}, {Vreeswijk}, {van Paradijs},
  {Kouveliotou}, {Augusteijn}, {B{\"o}hnhardt}, {Brewer}, {Doublier},
  {Gonzalez}, {Leibundgut}, {Lidman}, {Hainaut}, {Patat}, {Heise}, {in't Zand},
  {Hurley}, {Groot}, {Strom}, {Mazzali}, {Iwamoto}, {Nomoto}, {Umeda},
  {Nakamura}, {Young}, {Suzuki}, {Shigeyama}, {Koshut}, {Kippen}, {Robinson},
  {de Wildt}, {Wijers}, {Tanvir}, {Greiner}, {Pian}, {Palazzi}, {Frontera},
  {Masetti}, {Nicastro}, {Feroci}, {Costa}, {Piro}, {Peterson}, {Tinney},
  {Boyle}, {Cannon}, {Stathakis}, {Sadler}, {Begam}, \& {Ianna}}]{galama}
{Galama}, T.~J., {et~al.} 1998, \nat, 395, 670

\bibitem[{{Gall} {et~al.}(2011{\natexlab{a}}){Gall}, {Andersen}, \&
  {Hjorth}}]{gall10}
{Gall}, C., {Andersen}, A.~C., \& {Hjorth}, J. 2011{\natexlab{a}}, \aap, 528,
  A13

\bibitem[{{Gall} {et~al.}(2011{\natexlab{b}}){Gall}, {Andersen}, \&
  {Hjorth}}]{gall}
---. 2011{\natexlab{b}}, \aap, 528, A14

\bibitem[{{Gallerani} {et~al.}(2010){Gallerani}, {Maiolino}, {Juarez}, {Nagao},
  {Marconi}, {Bianchi}, {Schneider}, {Mannucci}, {Oliva}, {Willott}, {Jiang},
  \& {Fan}}]{gallerani2}
{Gallerani}, S., {et~al.} 2010, \aap, 523, A85

\bibitem[{{Gallerani} {et~al.}(2008){Gallerani}, {Salvaterra}, {Ferrara}, \&
  {Choudhury}}]{gallerani}
{Gallerani}, S., {Salvaterra}, R., {Ferrara}, A., \& {Choudhury}, T.~R. 2008,
  \mnras, 388, L84

\bibitem[{{Gehrels} {et~al.}(2004){Gehrels}, {Chincarini}, {Giommi}, {Mason},
  {Nousek}, {Wells}, {White}, {Barthelmy}, {Burrows}, {Cominsky}, {Hurley},
  {Marshall}, {M{\'e}sz{\'a}ros}, {Roming}, {Angelini}, {Barbier}, {Belloni},
  {Campana}, {Caraveo}, {Chester}, {Citterio}, {Cline}, {Cropper}, {Cummings},
  {Dean}, {Feigelson}, {Fenimore}, {Frail}, {Fruchter}, {Garmire}, {Gendreau},
  {Ghisellini}, {Greiner}, {Hill}, {Hunsberger}, {Krimm}, {Kulkarni}, {Kumar},
  {Lebrun}, {Lloyd-Ronning}, {Markwardt}, {Mattson}, {Mushotzky}, {Norris},
  {Osborne}, {Paczynski}, {Palmer}, {Park}, {Parsons}, {Paul}, {Rees},
  {Reynolds}, {Rhoads}, {Sasseen}, {Schaefer}, {Short}, {Smale}, {Smith},
  {Stella}, {Tagliaferri}, {Takahashi}, {Tashiro}, {Townsley}, {Tueller},
  {Turner}, {Vietri}, {Voges}, {Ward}, {Willingale}, {Zerbi}, \&
  {Zhang}}]{gehrels}
{Gehrels}, N., {et~al.} 2004, \apj, 611, 1005

\bibitem[{{Gehrz}(1989)}]{gehrz}
{Gehrz}, R. 1989, in IAU Symposium, Vol. 135, Interstellar Dust, ed. L.~J.
  {Allamandola} \& A.~G.~G.~M. {Tielens}, 445

\bibitem[{{Gendre} {et~al.}(2006){Gendre}, {Corsi}, \& {Piro}}]{gendre}
{Gendre}, B., {Corsi}, A., \& {Piro}, L. 2006, \aap, 455, 803

\bibitem[{{Gonz{\'a}lez} {et~al.}(2010){Gonz{\'a}lez}, {Labb{\'e}}, {Bouwens},
  {Illingworth}, {Franx}, {Kriek}, \& {Brammer}}]{valentino}
{Gonz{\'a}lez}, V., {Labb{\'e}}, I., {Bouwens}, R.~J., {Illingworth}, G.,
  {Franx}, M., {Kriek}, M., \& {Brammer}, G.~B. 2010, \apj, 713, 115

\bibitem[{{Granot} \& {Sari}(2002)}]{granot}
{Granot}, J., \& {Sari}, R. 2002, \apj, 568, 820

\bibitem[{{Greiner} {et~al.}(2009){Greiner}, {Kr{\"u}hler}, {Fynbo}, {Rossi},
  {Schwarz}, {Klose}, {Savaglio}, {Tanvir}, {McBreen}, {Totani}, {Zhang}, {Wu},
  {Watson}, {Barthelmy}, {Beardmore}, {Ferrero}, {Gehrels}, {Kann}, {Kawai},
  {Yolda{\c s}}, {M{\'e}sz{\'a}ros}, {Milvang-Jensen}, {Oates}, {Pierini},
  {Schady}, {Toma}, {Vreeswijk}, {Yolda{\c s}}, {Zhang}, {Afonso}, {Aoki},
  {Burrows}, {Clemens}, {Filgas}, {Haiman}, {Hartmann}, {Hasinger}, {Hjorth},
  {Jehin}, {Levan}, {Liang}, {Malesani}, {Pyo}, {Schulze}, {Szokoly}, {Terada},
  \& {Wiersema}}]{greiner}
{Greiner}, J., {et~al.} 2009, \apj, 693, 1610

\bibitem[{{Haislip} {et~al.}(2006){Haislip}, {Nysewander}, {Reichart}, {Levan},
  {Tanvir}, {Cenko}, {Fox}, {Price}, {Castro-Tirado}, {Gorosabel}, {Evans},
  {Figueredo}, {MacLeod}, {Kirschbrown}, {Jelinek}, {Guziy}, {Postigo},
  {Cypriano}, {Lacluyze}, {Graham}, {Priddey}, {Chapman}, {Rhoads}, {Fruchter},
  {Lamb}, {Kouveliotou}, {Wijers}, {Bayliss}, {Schmidt}, {Soderberg},
  {Kulkarni}, {Harrison}, {Moon}, {Gal-Yam}, {Kasliwal}, {Hudec}, {Vitek},
  {Kubanek}, {Crain}, {Foster}, {Clemens}, {Bartelme}, {Canterna}, {Hartmann},
  {Henden}, {Klose}, {Park}, {Williams}, {Rol}, {O'Brien}, {Bersier}, {Prada},
  {Pizarro}, {Maturana}, {Ugarte}, {Alvarez}, {Fernandez}, {Jarvis}, {Moles},
  {Alfaro}, {Ivarsen}, {Kumar}, {Mack}, {Zdarowicz}, {Gehrels}, {Barthelmy}, \&
  {Burrows}}]{haislip}
{Haislip}, J.~B., {et~al.} 2006, \nat, 440, 181

\bibitem[{{Hirashita} {et~al.}(2005){Hirashita}, {Nozawa}, {Kozasa}, {Ishii},
  \& {Takeuchi}}]{hirashita}
{Hirashita}, H., {Nozawa}, T., {Kozasa}, T., {Ishii}, T.~T., \& {Takeuchi},
  T.~T. 2005, \mnras, 357, 1077

\bibitem[{{Hjorth} {et~al.}(2003){Hjorth}, {Sollerman}, {M{\o}ller}, {Fynbo},
  {Woosley}, {Kouveliotou}, {Tanvir}, {Greiner}, {Andersen}, {Castro-Tirado},
  {Castro Cer{\'o}n}, {Fruchter}, {Gorosabel}, {Jakobsson}, {Kaper}, {Klose},
  {Masetti}, {Pedersen}, {Pedersen}, {Pian}, {Palazzi}, {Rhoads}, {Rol}, {van
  den Heuvel}, {Vreeswijk}, {Watson}, \& {Wijers}}]{hjorth}
{Hjorth}, J., {et~al.} 2003, \nat, 423, 847

\bibitem[{{Jakobsson} {et~al.}(2006){Jakobsson}, {Levan}, {Fynbo}, {Priddey},
  {Hjorth}, {Tanvir}, {Watson}, {Jensen}, {Sollerman}, {Natarajan},
  {Gorosabel}, {Castro Cer{\'o}n}, {Pedersen}, {Pursimo}, {{\'A}rnad{\'o}ttir},
  {Castro-Tirado}, {Davis}, {Deeg}, {Fiuza}, {Mykolaitis}, \&
  {Sousa}}]{jakobsson06}
{Jakobsson}, P., {et~al.} 2006, \aap, 447, 897

\bibitem[{{Kalberla} {et~al.}(2005){Kalberla}, {Burton}, {Hartmann}, {Arnal},
  {Bajaja}, {Morras}, \& {P{\"o}ppel}}]{kalberla}
{Kalberla}, P.~M.~W., {Burton}, W.~B., {Hartmann}, D., {Arnal}, E.~M.,
  {Bajaja}, E., {Morras}, R., \& {P{\"o}ppel}, W.~G.~L. 2005, \aap, 440, 775

\bibitem[{{Kann} {et~al.}(2010){Kann}, {Klose}, {Zhang}, {Malesani}, {Nakar},
  {Pozanenko}, {Wilson}, {Butler}, {Jakobsson}, {Schulze}, {Andreev},
  {Antonelli}, {Bikmaev}, {Biryukov}, {B{\"o}ttcher}, {Burenin}, {Castro
  Cer{\'o}n}, {Castro-Tirado}, {Chincarini}, {Cobb}, {Covino}, {D'Avanzo},
  {D'Elia}, {Della Valle}, {de Ugarte Postigo}, {Efimov}, {Ferrero}, {Fugazza},
  {Fynbo}, {G{\aa}lfalk}, {Grundahl}, {Gorosabel}, {Gupta}, {Guziy}, {Hafizov},
  {Hjorth}, {Holhjem}, {Ibrahimov}, {Im}, {Israel}, {Je{\'l}inek}, {Jensen},
  {Karimov}, {Khamitov}, {Kizilo{\v g}lu}, {Klunko}, {Kub{\'a}nek}, {Kutyrev},
  {Laursen}, {Levan}, {Mannucci}, {Martin}, {Mescheryakov}, {Mirabal},
  {Norris}, {Ovaldsen}, {Paraficz}, {Pavlenko}, {Piranomonte}, {Rossi},
  {Rumyantsev}, {Salinas}, {Sergeev}, {Sharapov}, {Sollerman}, {Stecklum},
  {Stella}, {Tagliaferri}, {Tanvir}, {Telting}, {Testa}, {Updike}, {Volnova},
  {Watson}, {Wiersema}, \& {Xu}}]{kann10}
{Kann}, D.~A., {et~al.} 2010, \apj, 720, 1513

\bibitem[{{Kawai} {et~al.}(2006){Kawai}, {Kosugi}, {Aoki}, {Yamada}, {Totani},
  {Ohta}, {Iye}, {Hattori}, {Aoki}, {Furusawa}, {Hurley}, {Kawabata},
  {Kobayashi}, {Komiyama}, {Mizumoto}, {Nomoto}, {Noumaru}, {Ogasawara},
  {Sato}, {Sekiguchi}, {Shirasaki}, {Suzuki}, {Takata}, {Tamagawa}, {Terada},
  {Watanabe}, {Yatsu}, \& {Yoshida}}]{kawai}
{Kawai}, N., {et~al.} 2006, \nat, 440, 184

\bibitem[{{Kozasa} {et~al.}(1991){Kozasa}, {Hasegawa}, \& {Nomoto}}]{kozasa}
{Kozasa}, T., {Hasegawa}, H., \& {Nomoto}, K. 1991, \aap, 249, 474

\bibitem[{{Lai} {et~al.}(2007){Lai}, {Huang}, {Fazio}, {Cowie}, {Hu}, \&
  {Kakazu}}]{lai}
{Lai}, K., {Huang}, J., {Fazio}, G., {Cowie}, L.~L., {Hu}, E.~M., \& {Kakazu},
  Y. 2007, \apj, 655, 704

\bibitem[{{Le Floc'h} {et~al.}(2003){Le Floc'h}, {Duc}, {Mirabel}, {Sanders},
  {Bosch}, {Diaz}, {Donzelli}, {Rodrigues}, {Courvoisier}, {Greiner},
  {Mereghetti}, {Melnick}, {Maza}, \& {Minniti}}]{lefloch}
{Le Floc'h}, E., {et~al.} 2003, \aap, 400, 499

\bibitem[{{Maiolino} {et~al.}(2004){Maiolino}, {Schneider}, {Oliva}, {Bianchi},
  {Ferrara}, {Mannucci}, {Pedani}, \& {Roca Sogorb}}]{maiolino}
{Maiolino}, R., {Schneider}, R., {Oliva}, E., {Bianchi}, S., {Ferrara}, A.,
  {Mannucci}, F., {Pedani}, M., \& {Roca Sogorb}, M. 2004, \nat, 431, 533

\bibitem[{{Malesani} {et~al.}(2004){Malesani}, {Tagliaferri}, {Chincarini},
  {Covino}, {Della Valle}, {Fugazza}, {Mazzali}, {Zerbi}, {D'Avanzo},
  {Kalogerakos}, {Simoncelli}, {Antonelli}, {Burderi}, {Campana}, {Cucchiara},
  {Fiore}, {Ghirlanda}, {Goldoni}, {G{\"o}tz}, {Mereghetti}, {Mirabel},
  {Romano}, {Stella}, {Minezaki}, {Yoshii}, \& {Nomoto}}]{malesani}
{Malesani}, D., {et~al.} 2004, \apjl, 609, L5

\bibitem[{{Marchenko}(2006)}]{marchenko}
{Marchenko}, S.~V. 2006, in Astronomical Society of the Pacific Conference
  Series, Vol. 353, Stellar Evolution at Low Metallicity: Mass Loss,
  Explosions, Cosmology, ed. {H.~J.~G.~L.~M.~Lamers, N.~Langer, T.~Nugis, \&
  K.~Annuk}, 299

\bibitem[{{Meikle} {et~al.}(2007){Meikle}, {Mattila}, {Pastorello}, {Gerardy},
  {Kotak}, {Sollerman}, {Van Dyk}, {Farrah}, {Filippenko}, {H{\"o}flich},
  {Lundqvist}, {Pozzo}, \& {Wheeler}}]{meikle}
{Meikle}, W.~P.~S., {et~al.} 2007, \apj, 665, 608

\bibitem[{{Micha{\l}owski} {et~al.}(2010){Micha{\l}owski}, {Murphy}, {Hjorth},
  {Watson}, {Gall}, \& {Dunlop}}]{michalowski}
{Micha{\l}owski}, M.~J., {Murphy}, E.~J., {Hjorth}, J., {Watson}, D., {Gall},
  C., \& {Dunlop}, J.~S. 2010, \aap, 522, A15

\bibitem[{{Morgan} \& {Edmunds}(2003)}]{morgan}
{Morgan}, H.~L., \& {Edmunds}, M.~G. 2003, \mnras, 343, 427

\bibitem[{{Nozawa} {et~al.}(2003){Nozawa}, {Kozasa}, {Umeda}, {Maeda}, \&
  {Nomoto}}]{nozawa}
{Nozawa}, T., {Kozasa}, T., {Umeda}, H., {Maeda}, K., \& {Nomoto}, K. 2003,
  \apj, 598, 785

\bibitem[{{Patel} {et~al.}(2010){Patel}, {Warren}, {Mortlock}, \&
  {Fynbo}}]{patel}
{Patel}, M., {Warren}, S.~J., {Mortlock}, D.~J., \& {Fynbo}, J.~P.~U. 2010,
  \aap, 512, L3

\bibitem[{{Perley} {et~al.}(2010){Perley}, {Bloom}, {Klein}, {Covino},
  {Minezaki}, {Wo{\'z}niak}, {Vestrand}, {Williams}, {Milne}, {Butler},
  {Updike}, {Kr{\"u}hler}, {Afonso}, {Antonelli}, {Cowie}, {Ferrero},
  {Greiner}, {Hartmann}, {Kakazu}, {K{\"u}pc{\"u} Yolda{\c s}}, {Morgan},
  {Price}, {Prochaska}, \& {Yoshii}}]{perley}
{Perley}, D.~A., {et~al.} 2010, \mnras, 406, 2473

\bibitem[{{Perna} \& {Lazzati}(2002)}]{perna02}
{Perna}, R., \& {Lazzati}, D. 2002, \apj, 580, 261

\bibitem[{{Pipino} {et~al.}(2011){Pipino}, {Fan}, {Matteucci}, {Calura},
  {Silva}, {Granato}, \& {Maiolino}}]{pipino}
{Pipino}, A., {Fan}, X.~L., {Matteucci}, F., {Calura}, F., {Silva}, L.,
  {Granato}, G., \& {Maiolino}, R. 2011, \aap, 525, A61

\bibitem[{{Prochaska} {et~al.}(2004){Prochaska}, {Bloom}, {Chen}, {Hurley},
  {Melbourne}, {Dressler}, {Graham}, {Osip}, \& {Vacca}}]{prochaska}
{Prochaska}, J.~X., {et~al.} 2004, \apj, 611, 200

\bibitem[{{Ruiz-Velasco} {et~al.}(2007){Ruiz-Velasco}, {Swan}, {Troja},
  {Malesani}, {Fynbo}, {Starling}, {Xu}, {Aharonian}, {Akerlof}, {Andersen},
  {Ashley}, {Barthelmy}, {Bersier}, {Castro Cer{\'o}n}, {Castro-Tirado},
  {Gehrels}, {G{\"o}{\u g}{\"u}{\c s}}, {Gorosabel}, {Guidorzi}, {G{\"u}ver},
  {Hjorth}, {Horns}, {Huang}, {Jakobsson}, {Jensen}, {K{\i}z{\i}lo{\u g}lu},
  {Kouveliotou}, {Krimm}, {Ledoux}, {Levan}, {Marsh}, {McKay}, {Melandri},
  {Milvang-Jensen}, {Mundell}, {O'Brien}, {{\"O}zel}, {Phillips}, {Quimby},
  {Rowell}, {Rujopakarn}, {Rykoff}, {Schaefer}, {Sollerman}, {Tanvir},
  {Th{\"o}ne}, {Urata}, {Vestrand}, {Vreeswijk}, {Watson}, {Wheeler}, {Wijers},
  {Wren}, {Yost}, {Yuan}, {Zhai}, \& {Zheng}}]{ruiz07}
{Ruiz-Velasco}, A.~E., {et~al.} 2007, \apj, 669, 1

\bibitem[{{Salvaterra} {et~al.}(2009){Salvaterra}, {Della Valle}, {Campana},
  {Chincarini}, {Covino}, {D'Avanzo}, {Fern{\'a}ndez-Soto}, {Guidorzi},
  {Mannucci}, {Margutti}, {Th{\"o}ne}, {Antonelli}, {Barthelmy}, {de Pasquale},
  {D'Elia}, {Fiore}, {Fugazza}, {Hunt}, {Maiorano}, {Marinoni}, {Marshall},
  {Molinari}, {Nousek}, {Pian}, {Racusin}, {Stella}, {Amati}, {Andreuzzi},
  {Cusumano}, {Fenimore}, {Ferrero}, {Giommi}, {Guetta}, {Holland}, {Hurley},
  {Israel}, {Mao}, {Markwardt}, {Masetti}, {Pagani}, {Palazzi}, {Palmer},
  {Piranomonte}, {Tagliaferri}, \& {Testa}}]{salvaterra}
{Salvaterra}, R., {et~al.} 2009, \nat, 461, 1258

\bibitem[{{Sari} {et~al.}(1998){Sari}, {Piran}, \& {Narayan}}]{sari}
{Sari}, R., {Piran}, T., \& {Narayan}, R. 1998, \apjl, 497, L17

\bibitem[{{Savaglio} {et~al.}(2009){Savaglio}, {Glazebrook}, \& {Le
  Borgne}}]{savaglio09}
{Savaglio}, S., {Glazebrook}, K., \& {Le Borgne}, D. 2009, \apj, 691, 182

\bibitem[{{Schady} {et~al.}(2010){Schady}, {Page}, {Oates}, {Still}, {de
  Pasquale}, {Dwelly}, {Kuin}, {Holland}, {Marshall}, \& {Roming}}]{schady}
{Schady}, P., {et~al.} 2010, \mnras, 401, 2773

\bibitem[{{Schaerer} \& {de Barros}(2010)}]{schaerer}
{Schaerer}, D., \& {de Barros}, S. 2010, \aap, 515, A73

\bibitem[{{Schlafly} {et~al.}(2010){Schlafly}, {Finkbeiner}, {Schlegel},
  {Juri{\'c}}, {Ivezi{\'c}}, {Gibson}, {Knapp}, \& {Weaver}}]{schlafly}
{Schlafly}, E.~F., {Finkbeiner}, D.~P., {Schlegel}, D.~J., {Juri{\'c}}, M.,
  {Ivezi{\'c}}, {\v Z}., {Gibson}, R.~R., {Knapp}, G.~R., \& {Weaver}, B.~A.
  2010, \apj, 725, 1175

\bibitem[{{Schlegel} {et~al.}(1998){Schlegel}, {Finkbeiner}, \&
  {Davis}}]{schlegel}
{Schlegel}, D.~J., {Finkbeiner}, D.~P., \& {Davis}, M. 1998, \apj, 500, 525

\bibitem[{{Skrutskie} {et~al.}(2006){Skrutskie}, {Cutri}, {Stiening},
  {Weinberg}, {Schneider}, {Carpenter}, {Beichman}, {Capps}, {Chester},
  {Elias}, {Huchra}, {Liebert}, {Lonsdale}, {Monet}, {Price}, {Seitzer},
  {Jarrett}, {Kirkpatrick}, {Gizis}, {Howard}, {Evans}, {Fowler}, {Fullmer},
  {Hurt}, {Light}, {Kopan}, {Marsh}, {McCallon}, {Tam}, {Van Dyk}, \&
  {Wheelock}}]{skrutskie}
{Skrutskie}, M.~F., {et~al.} 2006, \aj, 131, 1163

\bibitem[{{Smith} {et~al.}(2008){Smith}, {Foley}, \& {Filippenko}}]{smith}
{Smith}, N., {Foley}, R.~J., \& {Filippenko}, A.~V. 2008, \apj, 680, 568

\bibitem[{{Stanek} {et~al.}(2003){Stanek}, {Matheson}, {Garnavich}, {Martini},
  {Berlind}, {Caldwell}, {Challis}, {Brown}, {Schild}, {Krisciunas}, {Calkins},
  {Lee}, {Hathi}, {Jansen}, {Windhorst}, {Echevarria}, {Eisenstein}, {Pindor},
  {Olszewski}, {Harding}, {Holland}, \& {Bersier}}]{stanek}
{Stanek}, K.~Z., {et~al.} 2003, \apjl, 591, L17

\bibitem[{{Tagliaferri} {et~al.}(2005){Tagliaferri}, {Antonelli}, {Chincarini},
  {Fern{\'a}ndez-Soto}, {Malesani}, {Della Valle}, {D'Avanzo}, {Grazian},
  {Testa}, {Campana}, {Covino}, {Fiore}, {Stella}, {Castro-Tirado},
  {Gorosabel}, {Burrows}, {Capalbi}, {Cusumano}, {Conciatore}, {D'Elia},
  {Filliatre}, {Fugazza}, {Gehrels}, {Goldoni}, {Guetta}, {Guziy}, {Held},
  {Hurley}, {Israel}, {Jel{\'{\i}}nek}, {Lazzati}, {L{\'o}pez-Echarri},
  {Melandri}, {Mirabel}, {Moles}, {Moretti}, {Mason}, {Nousek}, {Osborne},
  {Pellizza}, {Perna}, {Piranomonte}, {Piro}, {de Ugarte Postigo}, \&
  {Romano}}]{tagliaferri}
{Tagliaferri}, G., {et~al.} 2005, \aap, 443, L1

\bibitem[{{Tanvir} {et~al.}(2009){Tanvir}, {Fox}, {Levan}, {Berger},
  {Wiersema}, {Fynbo}, {Cucchiara}, {Kr{\"u}hler}, {Gehrels}, {Bloom},
  {Greiner}, {Evans}, {Rol}, {Olivares}, {Hjorth}, {Jakobsson}, {Farihi},
  {Willingale}, {Starling}, {Cenko}, {Perley}, {Maund}, {Duke}, {Wijers},
  {Adamson}, {Allan}, {Bremer}, {Burrows}, {Castro-Tirado}, {Cavanagh}, {de
  Ugarte Postigo}, {Dopita}, {Fatkhullin}, {Fruchter}, {Foley}, {Gorosabel},
  {Kennea}, {Kerr}, {Klose}, {Krimm}, {Komarova}, {Kulkarni}, {Moskvitin},
  {Mundell}, {Naylor}, {Page}, {Penprase}, {Perri}, {Podsiadlowski}, {Roth},
  {Rutledge}, {Sakamoto}, {Schady}, {Schmidt}, {Soderberg}, {Sollerman},
  {Stephens}, {Stratta}, {Ukwatta}, {Watson}, {Westra}, {Wold}, \&
  {Wolf}}]{tanvir}
{Tanvir}, N.~R., {et~al.} 2009, \nat, 461, 1254

\bibitem[{{Todini} \& {Ferrara}(2001)}]{todini}
{Todini}, P., \& {Ferrara}, A. 2001, \mnras, 325, 726

\bibitem[{{Valiante} {et~al.}(2009){Valiante}, {Schneider}, {Bianchi}, \&
  {Andersen}}]{valiante}
{Valiante}, R., {Schneider}, R., {Bianchi}, S., \& {Andersen}, A.~C. 2009,
  \mnras, 397, 1661

\bibitem[{{Verma} {et~al.}(2007){Verma}, {Lehnert}, {F{\"o}rster Schreiber},
  {Bremer}, \& {Douglas}}]{verma}
{Verma}, A., {Lehnert}, M.~D., {F{\"o}rster Schreiber}, N.~M., {Bremer}, M.~N.,
  \& {Douglas}, L. 2007, \mnras, 377, 1024

\bibitem[{{Wang} {et~al.}(2008){Wang}, {Carilli}, {Wagg}, {Bertoldi}, {Walter},
  {Menten}, {Omont}, {Cox}, {Strauss}, {Fan}, {Jiang}, \& {Schneider}}]{wang}
{Wang}, R., {et~al.} 2008, \apj, 687, 848

\bibitem[{{Watson} {et~al.}(2010){Watson}, {French}, {Christensen},
  {O'Halloran}, {Micha{\l}owski}, {Hjorth}, {Malesani}, {Fynbo}, {Gordon}, \&
  {Castro Cer{\'o}n}}]{watson10}
{Watson}, D., {et~al.} 2010, \apj, submitted, arXiv:1010.1783

\bibitem[{{Watson} {et~al.}(2006{\natexlab{a}}){Watson}, {Fynbo}, {Ledoux},
  {Vreeswijk}, {Hjorth}, {Smette}, {Andersen}, {Aoki}, {Augusteijn},
  {Beardmore}, {Bersier}, {Castro Cer{\'o}n}, {D'Avanzo}, {Diaz-Fraile},
  {Gorosabel}, {Hirst}, {Jakobsson}, {Jensen}, {Kawai}, {Kosugi}, {Laursen},
  {Levan}, {Masegosa}, {N{\"a}r{\"a}nen}, {Page}, {Pedersen}, {Pozanenko},
  {Reeves}, {Rumyantsev}, {Shahbaz}, {Sharapov}, {Sollerman}, {Starling},
  {Tanvir}, {Torstensson}, \& {Wiersema}}]{watson06}
---. 2006{\natexlab{a}}, \apj, 652, 1011

\bibitem[{{Watson} {et~al.}(2007){Watson}, {Hjorth}, {Fynbo}, {Jakobsson},
  {Foley}, {Sollerman}, \& {Wijers}}]{watson07}
{Watson}, D., {Hjorth}, J., {Fynbo}, J.~P.~U., {Jakobsson}, P., {Foley}, S.,
  {Sollerman}, J., \& {Wijers}, R.~A.~M.~J. 2007, \apjl, 660, L101

\bibitem[{{Watson} {et~al.}(2006{\natexlab{b}}){Watson}, {Reeves}, {Hjorth},
  {Fynbo}, {Jakobsson}, {Pedersen}, {Sollerman}, {Castro Cer{\'o}n}, {McBreen},
  \& {Foley}}]{watson062}
{Watson}, D., {et~al.} 2006{\natexlab{b}}, \apjl, 637, L69

\bibitem[{{Woosley}(1993)}]{woosley}
{Woosley}, S.~E. 1993, \apj, 405, 273

\bibitem[{{Zafar} {et~al.}(2011){Zafar}, {Watson}, {Fynbo}, {Malesani},
  {Jakobsson}, \& {de Ugarte Postigo}}]{zafar11}
{Zafar}, T., {Watson}, D., {Fynbo}, J.~P.~U., {Malesani}, D., {Jakobsson}, P.,
  \& {de Ugarte Postigo}, A. 2011, \aap, accepted (arXiv:1102.1469)

\bibitem[{{Zafar} {et~al.}(2010){Zafar}, {Watson}, {Malesani}, {Vreeswijk},
  {Fynbo}, {Hjorth}, {Levan}, \& {Micha{\l}owski}}]{zafar}
{Zafar}, T., {Watson}, D.~J., {Malesani}, D., {Vreeswijk}, P.~M., {Fynbo},
  J.~P.~U., {Hjorth}, J., {Levan}, A.~J., \& {Micha{\l}owski}, M.~J. 2010,
  \aap, 515, A94

\end{thebibliography}

\end{document}